\documentclass{elsart}
\usepackage{epsfig,amssymb}

\begin{document}
\newcommand{\beqn}{\begin{equation}}
\newcommand{\eeqn}{\end{equation}}
\newcommand{\ee}{e^+e^-}
\newcommand{\pipipi}{\pi^+\pi^-\pi^0}
\newcommand{\etag}{\eta\gamma}
\newcommand{\pig}{\pi^0\gamma}
\newcommand{\eeetag}{\ee\to\etag}
\newcommand{\eepig}{\ee\to\pig}
\newcommand{\rhop}{\rho\prime}
\newcommand{\tg}{\gamma\gamma}
\newcommand{\br}{{\mathcal B}}
\newcommand{\brhoetag}{3.21 \pm 1.39 \pm 0.20 }
\newcommand{\bomegaetag}{4.44^{+2.59}_{-1.83}\pm 0.28 }
\newcommand{\bphietag}{1.373\pm 0.014\pm 0.085}
\newcommand{\brhopig}{6.21^{+1.28}_{-1.18}\pm 0.39}
\newcommand{\bomegapig}{9.06 \pm 0.20 \pm 0.57}
\newcommand{\bphipig}{1.258 \pm 0.037 \pm 0.077 }
\newcommand{\bib}{\bibitem}
\begin{frontmatter}
\date{}

\title{\large \bf \boldmath 
Study of the Processes $\eeetag$, $\pig\to 3\gamma$ in the 
c.m. Energy Range 600--1380~MeV at CMD-2}
\author[BINP]{R.R.~Akhmetshin},
\author[BINP,NGU]{V.M.~Aulchenko},
\author[BINP]{V.Sh.~Banzarov},
\author[PITT]{A.~Baratt},
\author[BINP,NGU]{L.M.~Barkov},
\author[BINP]{N.S.~Bashtovoy},
\author[BINP,NGU]{A.E.~Bondar},
\author[BINP]{D.V.~Bondarev},
\author[BINP]{A.V.~Bragin},
\author[BINP,NGU]{S.I.~Eidelman},
\author[BINP]{D.A.~Epifanov},
\author[BINP,NGU]{G.V.~Fedotovitch},
\author[BINP]{N.I.~Gabyshev},
\author[BINP]{D.A.~Gorbachev},
\author[BINP]{A.A.~Grebeniuk}, 
\author[BINP]{D.N.~Grigoriev},
\author[BINP]{F.V.~Ignatov},
\author[BINP]{S.V.~Karpov},
\author[BINP,NGU]{V.F.~Kazanin},
\author[BINP,NGU]{B.I.~Khazin},
\author[BINP,NGU]{I.A.~Koop},
\author[BINP,NGU]{P.P.~Krokovny},
\author[BINP,NGU]{A.S.~Kuzmin},
\author[BINP]{Yu.E.~Lischenko},
\author[BINP]{I.B.~Logashenko},
\author[BINP]{P.A.~Lukin},
\author[BINP]{K.Yu.~Mikhailov},
\author[BINP,NGU]{A.I.~Milstein},
\author[BINP,NGU]{I.N.~Nesterenko},
\author[BINP]{V.S.~Okhapkin},
\author[BINP,NGU]{A.V.~Otboev},
\author[BINP]{A.S.~Popov},
\author[BINP]{S.I.~Redin},
\author[BOST]{B.L.~Roberts},
\author[BINP]{N.I.~Root},
\author[BINP]{A.A.~Ruban},
\author[BINP]{N.M.~Ryskulov},
\author[BINP]{A.G.~Shamov}, 
\author[BINP]{Yu.M.~Shatunov},
\author[BINP,NGU]{B.A.~Shwartz},
\author[BINP,NGU]{A.L.~Sibidanov},
\author[BINP]{V.A.~Sidorov}, 
\author[BINP]{A.N.~Skrinsky},
\author[BINP]{I.G.~Snopkov},
\author[BINP,NGU]{E.P.~Solodov},
\author[PITT]{J.A.~Thompson}\footnote{deceased}, 
\author[BINP]{A.A.~Valishev},
\author[BINP]{Yu.V.~Yudin},
\author[BINP,NGU]{A.S.~Zaitsev},
\author[BINP]{S.G.~Zverev}

\address[BINP]{Budker Institute of Nuclear Physics, 
  Novosibirsk, 630090, Russia}
\address[BOST]{Boston University, Boston, MA 02215, USA}
\address[NGU]{Novosibirsk State University, 
  Novosibirsk, 630090, Russia}
\address[PITT]{University of Pittsburgh, Pittsburgh, PA 15260, USA}
\address[YALE]{Yale University, New Haven, CT 06511, USA}
\newpage
\begin{abstract}
The processes $\eeetag$, $\pig\to 3\gamma$ have been studied 
in the c.m. energy range 600--1380~MeV with the CMD-2 detector. 
The following branching ratios have been determined:
\begin{eqnarray*}
\br(\rho^0\to\etag)&=&(\brhoetag) \cdot 10^{-4}, \\
\br(\omega\to\etag)&=&(\bomegaetag)\cdot 10^{-4} ,\\
\br(\phi\to\etag)&=&(\bphietag) \cdot 10^{-2}, \\
\br(\rho^0\to\pig)&=&(\brhopig) \cdot 10^{-4}, \\
\br(\omega\to\pig)&=&(\bomegapig)\cdot 10^{-2}, \\
\br(\phi\to\pig)&=&(\bphipig) \cdot 10^{-3}.
\end{eqnarray*}
\end{abstract}
\end{frontmatter}
\maketitle

\section{Introduction}
The magnetic dipole transitions of the light vector mesons
($\rho$, $\omega$ and $\phi$) to the $\pig$ and $\etag$ final states 
have traditionally provided a convenient laboratory for various tests 
of theoretical concepts, particularly the nonrelativistic quark model 
and Vector Dominance Model (VDM)~\cite{th1,th2}. There are ongoing 
discussions about mechanisms of SU(3) breaking, possible admixture 
of glue in mesons and the role of anomalies in radiative 
decays~\cite{th3,th4,th5,th6,th7,th8}.  Precise measurements of the cross
sections of $\ee$ annihilation into the $\pig$ and $\etag$ 
final states in the broad c.m.energy range are necessary 
for the problem of the muon anomaly~\cite{amm}. Radiative decays
to $\pig$ and $\etag$ can also provide important information on
the properties of the $\rho$, $\omega$ and $\phi$ excitations as well as 
on the existence of light hybrids between 1000 and 
2000~MeV~\cite{hyb1,hyb2}. \\  

Despite previous experimental efforts (cf. the detailed bibliography
in~\cite{pdg}), of these decays only 
$\omega \to \pig$ and $\phi \to \etag$ are rather well studied.
A three-photon final state is convenient  for the investigation
of the $\pig$ and $\etag$ final states since both $\pi^0$ and
$\eta$ readily decay into two photons. Measurements
of the branching ratios for corresponding decays of the $\rho$, $\omega$
and $\phi$ using the two-photon
decay mode have been performed at ND~\cite{nd84,nd89} and 
SND~\cite{snd00,snd03}, however, none of them covered the whole 
off-resonance energy range. \\
 
In this work we report on the measurement of the cross section
of the processes $\ee \to \pig$ and $\ee \to \etag$ 
in the three-photon final state in the c.m.energy range
600--1380~MeV using the data from the CMD-2 detector at the VEPP-2M $\ee$ 
collider.

\section{Experiment}

The general purpose detector CMD-2 has been described in 
detail elsewhere~\cite{cmddet}. Its tracking system consists of a 
cylindrical drift chamber (DC) and double-layer multiwire proportional 
Z-chamber, both also used for the trigger, and both inside a thin 
(0.38~X$_0$) superconducting solenoid with a field of 1~T. 
The barrel CsI calorimeter (BC) with a thickness of 8.1~X$_0$ placed
outside  the solenoid has energy resolution for photons of about
9\% in the energy range from 100 to 700~MeV. The angular resolution is 
of the order of 0.02 radians. The end-cap BGO calorimeter with a 
thickness of 13.4~X$_0$ placed inside the solenoid 
has energy and angular resolution varying from 9\% to 4\% and from 
0.03 to 0.02 radians, respectively, for the photon energy in the range 
100 to 700~MeV. The barrel and end-cap calorimeter systems cover a 
solid angle of $0.92\times4\pi$ radians. \\ 

This analysis is based on a data sample corresponding to
integrated luminosity of 21~pb$^{-1}$ collected in 1997--1998 
in the energy range 600--1380~MeV.  
The step of the c.m. energy scan varied from 0.5~MeV near the $\omega$
and $\phi$ peaks to 10~MeV far from the resonances.
The beam energy spread is about $4\times 10^{-4}$ of the total energy.
The luminosity is measured using events of Bhabha scattering 
at large angles~\cite{prep}. \\

A GEANT3 based Monte Carlo simulation (MC) package is used to model 
the detector response and determine the efficiency~\cite{mccmd}.
Because of the beam induced background additional (``fake'') clusters 
can appear in the calorimeter. To take this effect into account
in MC we determine a corresponding probability as well as photon energy 
and angular spectra directly from the data using the process 
$\ee\to \to \pipipi$, and then include generation of such 
photons in the detector response during simulation.

\section{Data analysis}

\begin{figure}
 \includegraphics[width=0.5\textwidth]
{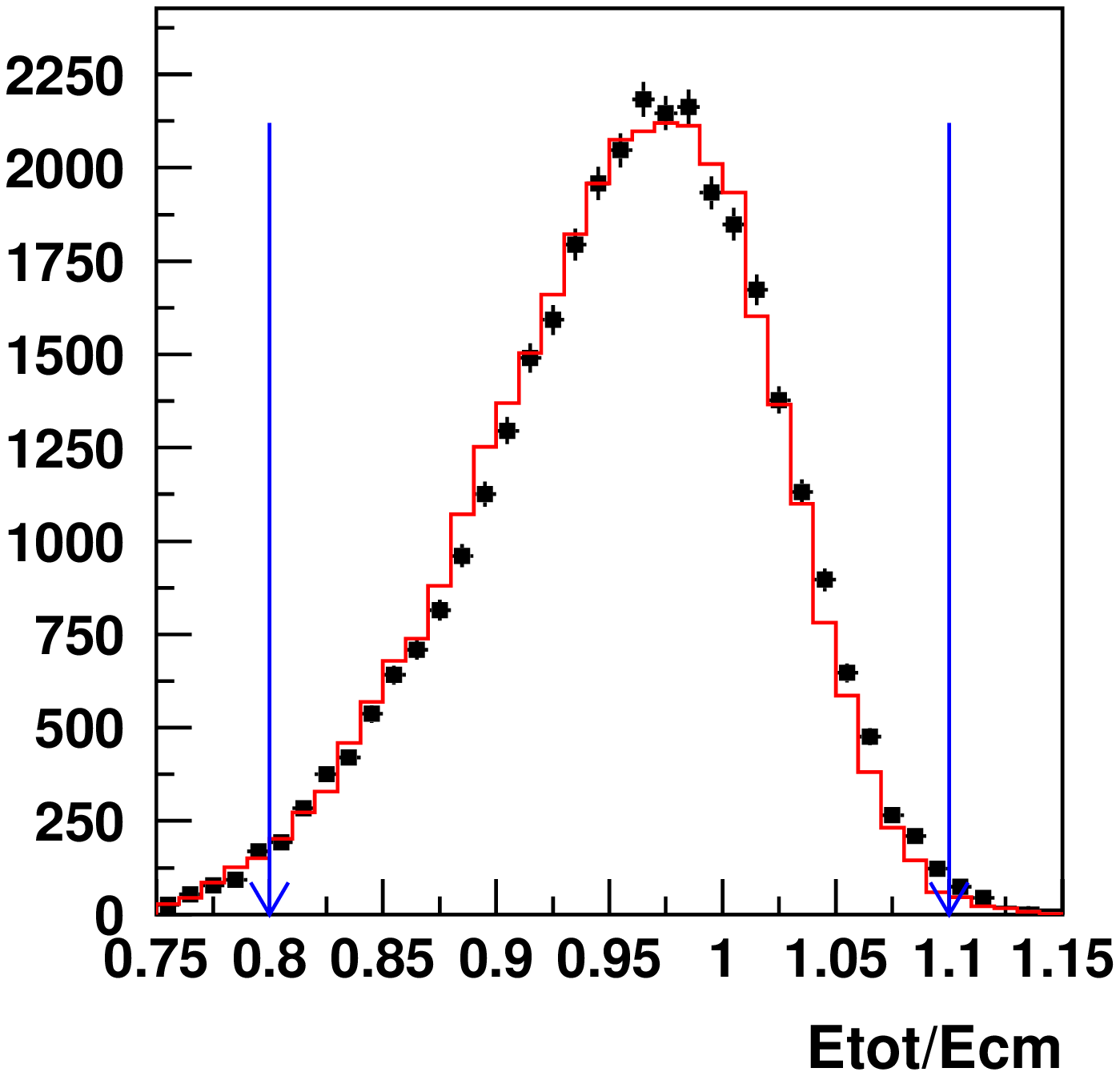}
 \includegraphics[width=0.5\textwidth]
{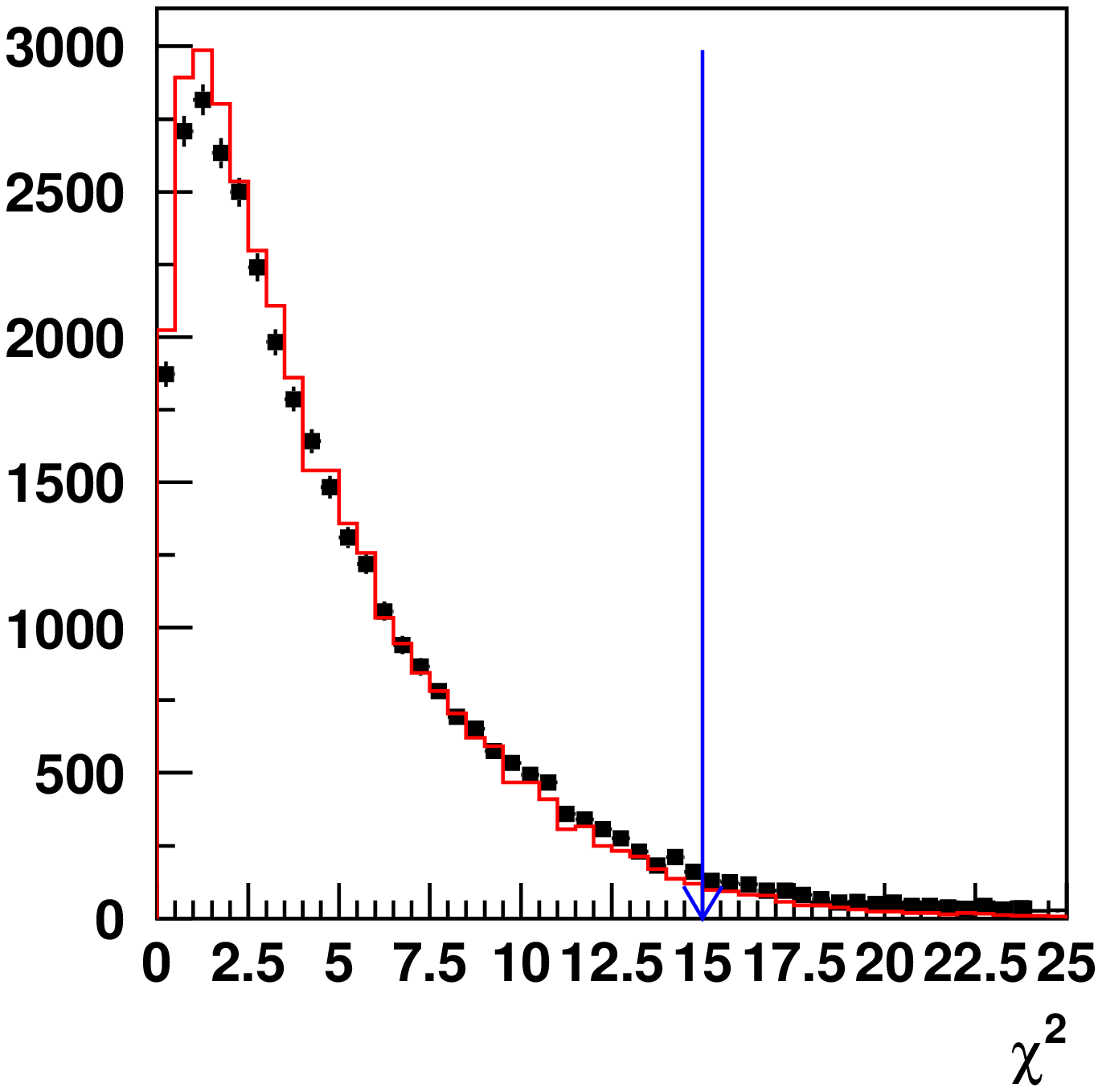}
  \caption{The $E_{\rm tot}/E_{\rm cm}$ (left) and $\chi^2$  (right) 
    distributions. The 
    points with error bars represent experimental events, the
    histograms show the MC simulation.
    The arrows indicate the cuts imposed. }
  \label{sel_comp}
\end{figure}

At the initial stage, events are selected which have no tracks in the DC, 
three or four photons in the CsI calorimeter, the total energy 
deposition $0.8 < E_{\rm tot}/ E_{\rm cm} < 1.1$, the total momentum 
$P_{\rm tot}/ E_{\rm cm}<0.15$ and the minimum photon energy of 50~MeV. 
Figure~\ref{sel_comp} (left) shows the $E_{\rm tot}$ distribution 
for the data and signal MC near the $\phi$ resonance. One can  
see good agreement between the data and signal MC.
About $52\times 10^3$ events were selected in the whole energy range
after these requirements.

Then a kinematic fit requiring energy-momentum conservation (a standard 4C fit) 
was performed. We require $\chi^2 < 15$ that provides a good signal/noise
ratio while the number of rejected signal events is still small
(see Fig.~\ref{sel_comp}).

The reconstruction procedure assumes three photons, i.e., for
events with four photons a combination of three photons with the
minimum $\chi^2$ is chosen.
After this stage about $48\times 10^3$ events remain.

The dominant background comes from the QED three-photon annihilation:
$\ee\to 3\gamma$. These events can not be completely rejected by
selection criteria. 
The $\etag$, $\pig$ and background events can be separated using
decay dynamics. To this end two methods were considered: Dalitz plot
analysis and a fit of the two-photon invariant mass distributions. 

\begin{figure}
\centering
 \includegraphics[width=0.6\textwidth]
{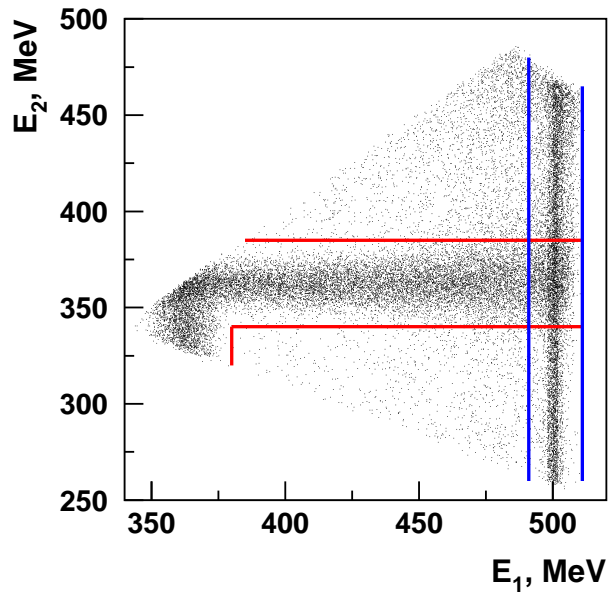}
 \caption{The Dalitz plot for the $3\gamma$ final state at the 
$\phi$ meson energy.
   The points represent experimental events,
   the lines indicate boundaries used in the selection criteria,
see the text for more detail.}
  \label{dalitz}
\end{figure}

Figure~\ref{dalitz} shows the Dalitz plot for the
$3\gamma$ final state in the $\phi$ meson energy range 
($1011.4$~MeV $< E_{\rm CM} < 1027.4$~MeV). 
Here the photons are sorted by their 
energy so that the first photon has a maximum energy: 
$E_{1} > E_{2} > E_{3}$.
The Dalitz plot is divided into three regions: $D_{\eta\gamma}$
({$340~{\rm MeV} < E_{2} < 385~{\rm MeV} 
~{\rm or}$} {$E_{1} < 385~{\rm MeV}$}),
$D_{\pi^0\gamma}$ ($491~{\rm MeV} < E_{1} < 511~{\rm MeV}$) and 
$D_{\rm bg}$ (all the remaining events).
For each of the three final states ($\etag$, $\pig$ and QED) we 
determine from the MC simulation the
probabilities to enter each region. Based on that, from the population
of various regions of the Dalitz plot in the data
the total number of events due to each process is calculated.
However, this method can provide bias in the signal yield determination 
because of the possible deviation between the signal shape in the data and
MC simulation. Additional bias can arise from the background processes
of the non-QED origin.

Therefore, we obtain the number of $\etag$ and $\pig$ events
by fitting the two-photon invariant mass distribution. In this method the 
signal shape is obtained from the data decreasing a possible bias.
The difference in the number of selected events in these two methods
(about 3\%) was considered as a systematic uncertainty because of
the separation procedure.

\begin{figure}
 \includegraphics[width=0.5\textwidth]
{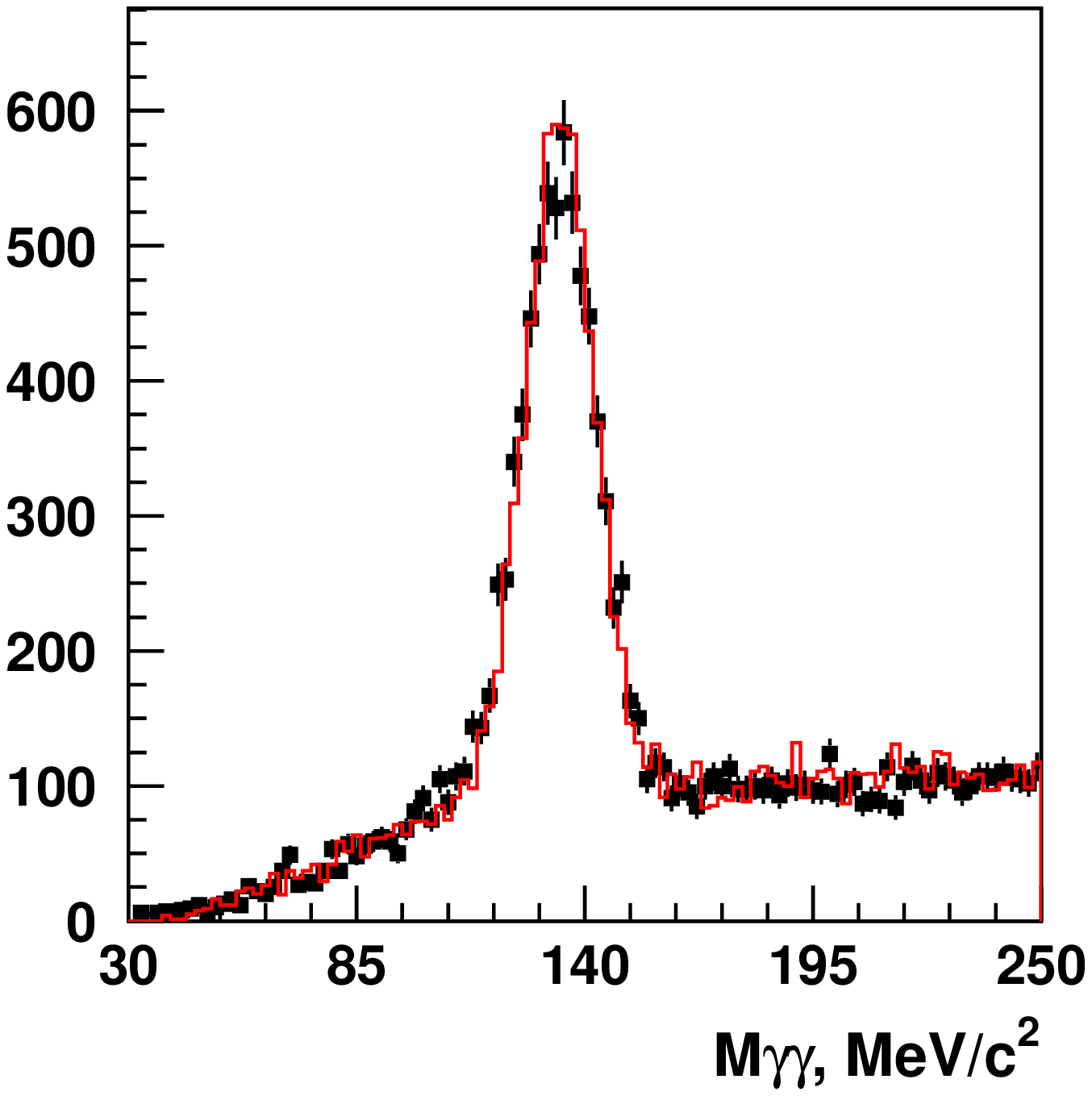}
 \includegraphics[width=0.5\textwidth]
{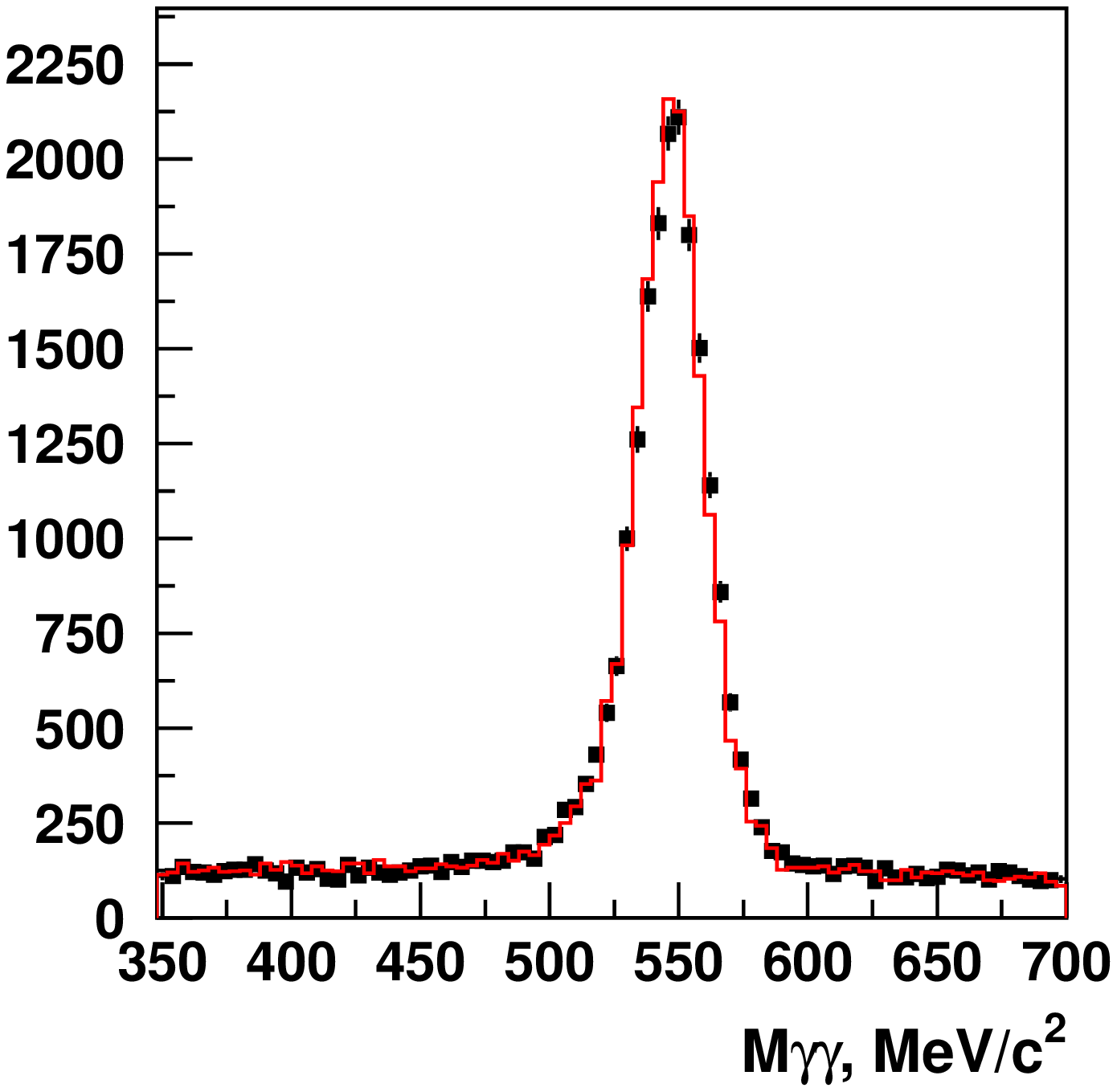}
  \caption{The two-photon invariant mass distributions in the 
  $\pi^0$ (left) and
   $\eta$ (right) mass range.
    The points with error bars represent experimental events, 
    histograms show the MC simulation.}
  \label{mass_comp}
\end{figure}

For $\pi^0$ reconstruction the invariant mass of the two softer photons 
($M_{23}$) is used.  For the $\eta$ signal three combinations are used:
\begin{enumerate}
\item In case of $E_1<m_\eta^2/\sqrt{s}$, two hard photons are used 
  ($M_{12}$).
\item Otherwise, if $E_3<m_\eta^2/\sqrt{s}$, we use the first and third 
  photons ($M_{13}$) .
\item In other cases two soft photons ($M_{23}$) are used.
\end{enumerate}
Figure~\ref{mass_comp} shows the two-photon invariant mass distributions
for the $\pi^0$ (left) and $\eta$ (right) combinations 
near the peak of the $\phi$ meson. 
  
Other possible sources of background are the processes 
$\ee \to \etag \to 3\pi^0\gamma$, 
$\ee \to K_{\rm S}K_{\rm L}$, $\ee \to \gamma\gamma$
and $\ee \to \omega \pi^0 \to \pi^0\pi^0\gamma$. The expected number
of events from these processes was calculated from the detection 
efficiencies determined by the MC simulation and their cross sections  
independently measured at CMD-2~\cite{cmdeta,cmdk1,cmdk2,cmdop1,cmdop2}.
The  fraction of background events is negligible below the $\phi$
meson and is  about 2\% only in the $\phi$ meson energy range.
Above the $\phi$ meson the expected cross section of the signal is
very low (0.01--0.1 nb) and that of the background remaining after all
selection criteria has close or even higher value. 
The separation procedure gives 17400 $\etag$
events, 18680 $\pig$ events and about 12000 QED events in the whole
energy range considered. 
 
\subsection{Approximation of the cross sections}

At each energy point $i$ the cross section of the process $\sigma_{\rm i}$ of
a given process is 
calculated using the following formula:
\begin{equation}
\sigma_{\rm i}=\frac{N_{\rm i}}
{L_{\rm i} \varepsilon_{\rm i} (1+\delta_{\rm i})}
\, , \label{Nth}
\end{equation}
where $N_{\rm i}$ is the number of selected events, 
$L_{\rm i}$ is the integrated luminosity, 
$\varepsilon_{\rm i}$ is the detection efficiency and
$(1+\delta_{\rm i})$ is the radiative correction at the i-th energy
point.  

The detection efficiency was calculated from the
Monte Carlo simulation 
taking into account corrections obtained from the data 
and  the neutral trigger efficiency. The neutral trigger (NT)
is part of the CMD-2 trigger system responsible for events
with a final state of photons only, without any charged tracks. 
The NT efficiency was estimated using events of
the process $\ee\to\ee\gamma$ at each energy point.
Its value varied from about 80\% to 90\%.

The radiative corrections are calculated according to~\cite{radcor}.
The dependence of the detection efficiency on the energy 
of the emitted photon is determined from simulation.

\begin{table*}
\caption{The c.m. energy, integrated luminosity, 
  number of selected events, detection efficiency, radiative 
  correction, and Born cross section $\sigma$ of the 
  process $\eeetag$.}
\medskip
\begin{tabular*}{\textwidth}{c@{\extracolsep{\fill}}rcrcr}
\hline\hline
$\sqrt{s}$, MeV & $L$, nb$^{-1}$ & $N_{\rm exp}$ & $\varepsilon$, \% &
$1+\delta$& $\sigma$, nb 
\\\hline
 599.86 &   35.2 & $    0.1\pm  2.3$ &  2.3 & -0.157 & $   <4.71$\\
 629.86 &   44.6 & $    0.7\pm  2.8$ &  7.2 & -0.142 & $   <1.68$\\
 659.86 &   39.8 & $    3.0\pm  2.9$ & 11.6 & -0.134 & $   0.75\pm 0.63$\\
 719.86 &   56.9 & $    0.0\pm  2.5$ & 15.0 & -0.126 & $   <0.48$\\
 749.86 &   42.9 & $    0.1\pm  1.9$ & 16.0 & -0.117 & $   <0.44$\\
 759.86 &   33.7 & $    3.4\pm  2.7$ & 15.7 & -0.115 & $   0.73\pm 0.51$\\
 763.86 &   39.7 & $    5.6\pm  3.9$ & 16.0 & -0.116 & $   1.00\pm 0.61$\\
 769.86 &   34.3 & $    0.2\pm  2.1$ & 17.0 & -0.126 & $   <0.63$\\
 773.86 &   70.1 & $    1.0\pm  2.8$ & 15.8 & -0.147 & $   0.11\pm 0.26$\\
 777.86 &   83.6 & $    2.4\pm  3.5$ & 15.8 & -0.186 & $   0.22\pm 0.27$\\
 779.86 &   56.6 & $   10.0\pm  3.9$ & 14.7 & -0.204 & $   1.50\pm 0.46$\\
 780.86 &   58.5 & $    2.3\pm  5.9$ &  8.8 & -0.207 & $   0.57\pm 1.14$\\
 781.86 &  366.8 & $   41.2\pm  9.3$ & 14.7 & -0.203 & $   0.94\pm 0.18$\\
 782.86 &   77.6 & $    9.8\pm  4.4$ & 15.3 & -0.191 & $   1.03\pm 0.37$\\
 783.86 &   71.7 & $    2.1\pm  3.3$ & 15.7 & -0.172 & $   0.22\pm 0.29$\\
 785.86 &   67.0 & $    8.2\pm  3.9$ & 16.0 & -0.123 & $   0.87\pm 0.36$\\
 789.86 &   28.4 & $    1.5\pm  2.1$ & 15.2 & -0.046 & $   0.36\pm 0.48$\\
 793.86 &   46.2 & $    1.7\pm  2.2$ & 15.9 & -0.010 & $   0.23\pm 0.31$\\
 799.86 &   56.5 & $    0.9\pm  2.0$ & 15.5 &  0.005 & $   0.10\pm 0.23$\\
 809.86 &   59.9 & $    3.9\pm  2.7$ & 15.9 &  0.006 & $   0.41\pm 0.29$\\
 819.86 &  109.4 & $    8.0\pm  4.1$ & 15.8 &  0.004 & $   0.47\pm 0.24$\\
 839.86 &  130.4 & $    8.6\pm  4.5$ & 15.4 & -0.006 & $   0.43\pm 0.23$\\
 879.86 &  167.9 & $    2.1\pm  3.9$ & 15.2 & -0.037 & $   0.08\pm 0.15$\\
 919.86 &  285.4 & $    5.7\pm  5.4$ & 15.0 & -0.063 & $   0.14\pm 0.13$\\
 939.86 &  136.7 & $    1.1\pm  3.8$ & 15.5 & -0.077 & $   0.06\pm 0.18$\\
 949.86 &  226.1 & $   12.5\pm  5.5$ & 16.2 & -0.085 & $   0.38\pm 0.15$\\
 957.86 &  250.1 & $    6.2\pm  4.7$ & 16.5 & -0.093 & $   0.17\pm 0.12$\\
 969.86 &  249.7 & $    4.7\pm  5.1$ & 17.2 & -0.108 & $   0.12\pm 0.12$\\
 983.93 &  307.7 & $    5.2\pm  7.0$ & 20.4 & -0.132 & $   0.07\pm 0.13$\\

\hline\hline
\end{tabular*}
\label{tab:etag_cs1}
\end{table*}

\begin{table*}
\caption{The c.m. energy, integrated luminosity, 
  number of selected events, detection efficiency, radiative 
  correction, and Born cross section $\sigma$ of the 
  process $\eeetag$.}
\medskip
\begin{tabular*}{\textwidth}{c@{\extracolsep{\fill}}rcrcr}
\hline\hline
$\sqrt{s}$, MeV & $L$, nb$^{-1}$ & $N_{\rm exp}$ & $\varepsilon$, \% &
$1+\delta$& $\sigma$, nb 
\\\hline
1003.91 &  357.7 & $   44.0\pm 10.2$ & 20.3 & -0.192 & $   0.67\pm 0.15$\\
1010.53 &  477.3 & $  109.6\pm 14.9$ & 19.9 & -0.227 & $   1.48\pm 0.16$\\
1015.77 &  391.7 & $  401.3\pm 23.0$ & 20.0 & -0.268 & $   6.80\pm 0.30$\\
1016.77 &  660.1 & $  968.1\pm 34.8$ & 19.6 & -0.277 & $  10.00\pm 0.27$\\
1016.91 &  306.1 & $  497.6\pm 24.9$ & 20.0 & -0.277 & $  10.90\pm 0.49$\\
1017.61 &  673.7 & $ 1362.1\pm 40.8$ & 20.1 & -0.282 & $  13.77\pm 0.32$\\
1017.77 &  563.1 & $ 1198.8\pm 38.2$ & 19.9 & -0.282 & $  14.80\pm 0.44$\\
1018.58 &  410.1 & $ 1230.8\pm 38.4$ & 20.1 & -0.278 & $  21.31\pm 0.58$\\
1018.83 &  977.5 & $ 2855.9\pm 57.8$ & 19.9 & -0.274 & $  21.13\pm 0.30$\\
1019.50 &  633.1 & $ 1941.9\pm 47.5$ & 20.1 & -0.254 & $  21.50\pm 0.43$\\
1019.84 &  810.8 & $ 2584.6\pm 54.6$ & 20.2 & -0.238 & $  21.54\pm 0.73$\\
1020.62 &  876.3 & $ 2231.7\pm 51.3$ & 20.0 & -0.187 & $  15.52\pm 0.30$\\
1021.54 &  440.6 & $  800.5\pm 30.9$ & 20.0 & -0.112 & $   9.82\pm 0.36$\\
1022.79 &  551.0 & $  621.5\pm 28.0$ & 20.1 &  0.007 & $   5.38\pm 0.26$\\
1027.67 &  562.2 & $  198.9\pm 17.3$ & 20.2 &  0.591 & $   1.08\pm 0.15$\\
1033.67 &  510.8 & $  100.0\pm 14.0$ & 19.9 &  1.557 & $   0.38\pm 0.14$\\
1039.59 &  447.5 & $   66.5\pm 11.9$ & 20.1 &  2.911 & $   0.18\pm 0.13$\\
1049.80 &  312.5 & $   23.4\pm  8.2$ & 19.6 &  6.778 & $   0.05\pm 0.13$\\
1059.49 &  220.6 & $    9.8\pm  5.9$ & 19.3 & 13.272 & $   <0.30$\\
1079.00 &  437.0 & $    4.6\pm  6.8$ & 22.7 & 39.899 & $   <0.11$\\
1163.40 &  918.2 & $    0.0\pm 10.3$ & 21.8 &  0.035 & $   <0.08$\\
1310.00 & 4249.0 & $   -0.4\pm 21.6$ & 20.9 & -0.074 & $   <0.05$\\

\hline\hline
\end{tabular*}
\label{tab:etag_cs2}
\end{table*}

\begin{table*}
\vspace*{-9mm}
\caption{The c.m. energy, integrated luminosity, 
  number of selected events, detection efficiency, radiative 
  correction, and Born cross section $\sigma$ of the 
  process $\eepig$.}
\medskip
\begin{tabular*}{\textwidth}{c@{\extracolsep{\fill}}rcrcr}
\hline\hline
$\sqrt{s}$, MeV & $L$, nb$^{-1}$& $N_{\rm exp}$ & $\varepsilon$, \% &
$1+\delta$& $\sigma$, nb 
\\\hline
 599.86 &   35.2 & $    4.8\pm  3.6$ & 12.0 & -0.089 & $   1.23\pm 0.86$\\
 629.86 &   44.6 & $    9.2\pm  4.2$ & 12.8 & -0.093 & $   1.78\pm 0.74$\\
 659.86 &   39.8 & $    8.6\pm  4.4$ & 12.5 & -0.099 & $   1.92\pm 0.89$\\
 719.86 &   56.9 & $   14.1\pm  5.2$ & 14.0 & -0.112 & $   2.00\pm 0.65$\\
 749.86 &   42.9 & $   27.1\pm  5.9$ & 14.3 & -0.131 & $   5.08\pm 0.97$\\
 759.86 &   33.7 & $   35.0\pm  6.8$ & 14.7 & -0.150 & $   8.31\pm 1.40$\\
 763.86 &   39.7 & $   62.8\pm  8.6$ & 14.5 & -0.162 & $  12.97\pm 1.52$\\
 769.86 &   34.3 & $   76.7\pm  9.2$ & 15.5 & -0.185 & $  17.64\pm 1.77$\\
 773.86 &   70.1 & $  281.1\pm 17.3$ & 14.6 & -0.204 & $  34.33\pm 1.82$\\
 777.86 &   83.6 & $  721.5\pm 27.5$ & 14.4 & -0.224 & $  76.65\pm 2.64$\\
 779.86 &   56.6 & $  757.7\pm 27.8$ & 13.8 & -0.229 & $ 125.81\pm 4.92$\\
 780.86 &   58.5 & $  717.7\pm 27.0$ &  8.6 & -0.228 & $ 184.98\pm10.99$\\
 781.86 &  366.8 & $ 6619.7\pm 82.0$ & 13.6 & -0.221 & $ 172.26\pm 2.47$\\
 782.86 &   77.6 & $ 1664.6\pm 41.1$ & 14.9 & -0.206 & $ 183.37\pm 4.80$\\
 783.86 &   71.7 & $ 1403.6\pm 37.7$ & 14.9 & -0.183 & $ 162.00\pm 4.72$\\
 785.86 &   67.0 & $  978.8\pm 31.6$ & 13.9 & -0.116 & $ 118.44\pm 4.16$\\
 789.86 &   28.4 & $  187.8\pm 13.9$ & 14.6 &  0.050 & $  42.80\pm 3.80$\\
 793.86 &   46.2 & $  166.2\pm 13.3$ & 14.8 &  0.217 & $  19.93\pm 2.06$\\
 799.86 &   56.5 & $  134.8\pm 12.2$ & 14.8 &  0.441 & $  11.18\pm 1.52$\\
 809.86 &   59.9 & $   83.7\pm  9.9$ & 15.5 &  0.724 & $   5.22\pm 1.09$\\
 819.86 &  109.4 & $   87.9\pm 10.4$ & 15.4 &  0.906 & $   2.74\pm 0.62$\\
 839.86 &  130.4 & $   61.7\pm  9.2$ & 15.8 &  0.901 & $   1.58\pm 0.45$\\
 879.86 &  167.9 & $   17.2\pm  6.0$ & 17.2 &  0.342 & $   0.44\pm 0.21$\\
 919.86 &  285.4 & $   20.8\pm  6.6$ & 17.4 &  0.021 & $   0.41\pm 0.13$\\
 939.86 &  136.7 & $   18.0\pm  5.5$ & 17.9 &  0.001 & $   0.74\pm 0.22$\\
 949.86 &  226.1 & $   20.1\pm  6.2$ & 18.0 & -0.008 & $   0.50\pm 0.15$\\
 957.86 &  250.1 & $   15.7\pm  5.8$ & 18.4 & -0.015 & $   0.35\pm 0.13$\\
 969.86 &  249.7 & $   11.8\pm  5.4$ & 18.6 & -0.029 & $   0.26\pm 0.12$\\
 983.93 &  307.7 & $    9.4\pm  6.3$ & 19.9 & -0.053 & $   0.16\pm 0.11$\\

\hline\hline
\end{tabular*}
\label{tab:pig_cs1}
\end{table*}

\begin{table*}
\vspace*{-9mm}
\caption{The c.m. energy, integrated luminosity, 
  number of selected events, detection efficiency, radiative 
  correction, and Born cross section $\sigma$ of the 
  process $\eepig$.}
\medskip
\begin{tabular*}{\textwidth}{c@{\extracolsep{\fill}}rcrcr}
\hline\hline
$\sqrt{s}$, MeV & $L$, nb$^{-1}$& $N_{\rm exp}$ & $\varepsilon$, \% &
$1+\delta$& $\sigma$, nb 
\\\hline
1003.91 &  357.7 & $   29.5\pm  8.2$ & 20.5 & -0.127 & $   0.44\pm 0.12$\\
1010.53 &  477.3 & $   50.3\pm 10.1$ & 20.7 & -0.179 & $   0.61\pm 0.10$\\
1015.77 &  391.7 & $  120.9\pm 13.3$ & 20.5 & -0.243 & $   1.95\pm 0.17$\\
1016.77 &  660.1 & $  306.1\pm 20.7$ & 20.4 & -0.256 & $   2.95\pm 0.16$\\
1016.91 &  306.1 & $  175.4\pm 15.3$ & 21.0 & -0.257 & $   3.59\pm 0.30$\\
1017.61 &  673.7 & $  401.4\pm 23.3$ & 20.3 & -0.263 & $   3.97\pm 0.18$\\
1017.77 &  563.1 & $  363.4\pm 22.0$ & 20.5 & -0.264 & $   4.29\pm 0.24$\\
1018.58 &  410.1 & $  347.8\pm 21.5$ & 20.7 & -0.260 & $   5.72\pm 0.32$\\
1018.83 &  977.5 & $  764.3\pm 32.1$ & 20.2 & -0.255 & $   5.46\pm 0.17$\\
1019.50 &  633.1 & $  466.7\pm 25.2$ & 20.7 & -0.228 & $   4.77\pm 0.24$\\
1019.84 &  810.8 & $  591.8\pm 28.5$ & 20.8 & -0.208 & $   4.54\pm 0.52$\\
1020.62 &  876.3 & $  454.3\pm 26.0$ & 20.7 & -0.139 & $   2.85\pm 0.15$\\
1021.54 &  440.6 & $  143.4\pm 15.2$ & 20.6 & -0.024 & $   1.52\pm 0.17$\\
1022.79 &  551.0 & $  114.6\pm 14.0$ & 20.4 &  0.199 & $   0.78\pm 0.13$\\
1027.67 &  562.2 & $   35.1\pm  9.2$ & 20.6 &  2.660 & $   0.08\pm 0.08$\\
1033.67 &  510.8 & $   13.8\pm  7.5$ & 20.7 & 43.316 & $   <0.11$\\
1039.59 &  447.5 & $   10.4\pm  6.6$ & 20.7 & 72.963 & $   <0.11$\\
1049.80 &  312.5 & $    1.8\pm  5.1$ & 20.4 &  6.939 & $   <0.13$\\
1059.49 &  220.6 & $    2.0\pm  5.3$ & 20.4 &  3.314 & $   <0.20$\\
1079.00 &  437.0 & $   -0.6\pm  5.5$ & 27.4 &  1.634 & $   <0.08$\\
1163.40 &  918.2 & $   19.8\pm  9.7$ & 28.1 & -0.047 & $   0.08\pm 0.04$\\
1310.00 & 4249.0 & $   48.1\pm 16.2$ & 26.9 & -0.143 & $   0.05\pm 0.02$\\

\hline\hline
\end{tabular*}
\label{tab:pig_cs2}
\end{table*}

\begin{figure*}
\centering
\includegraphics[width=0.5\textwidth]
{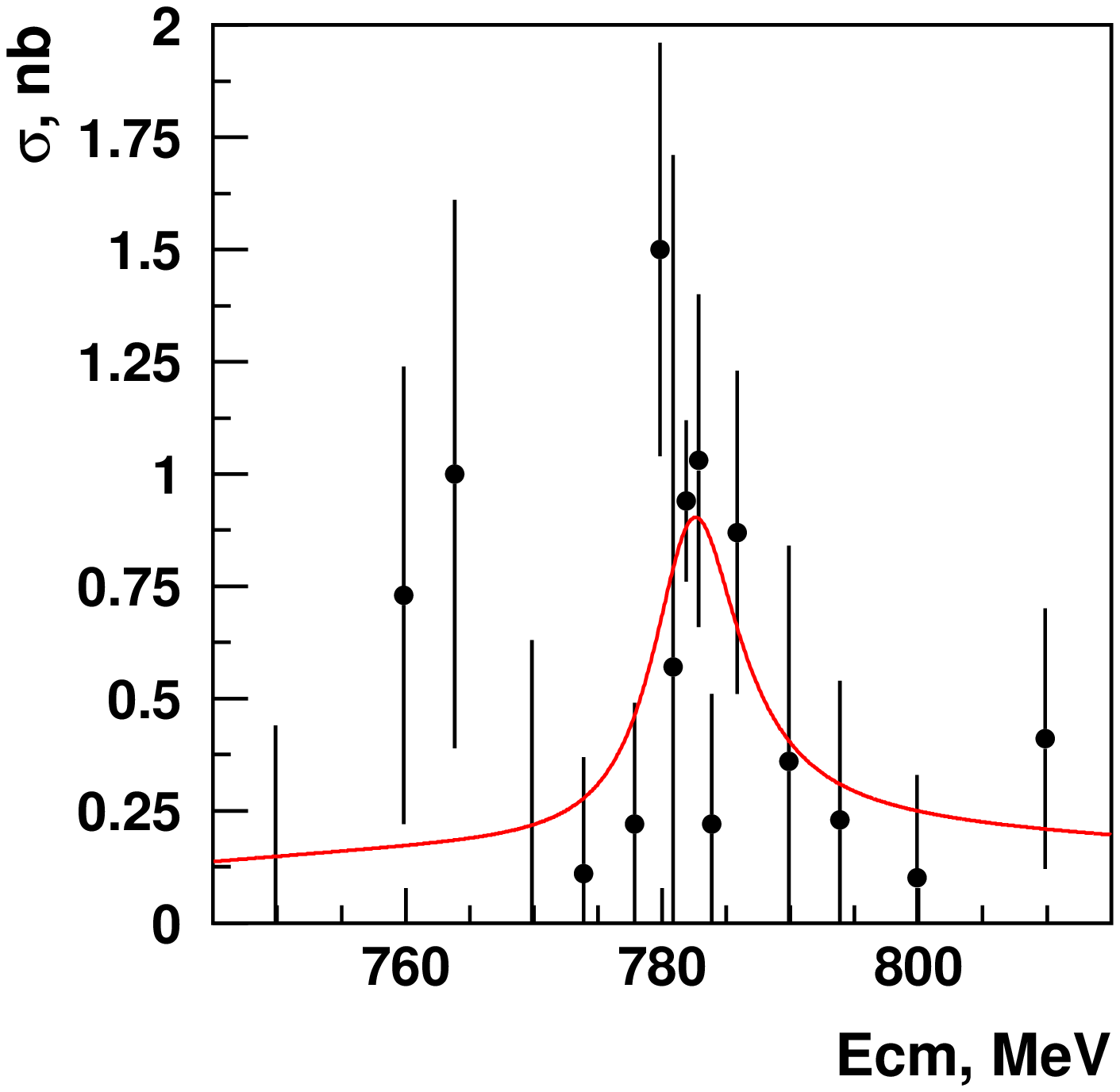}\hfill
\includegraphics[width=0.5\textwidth]
{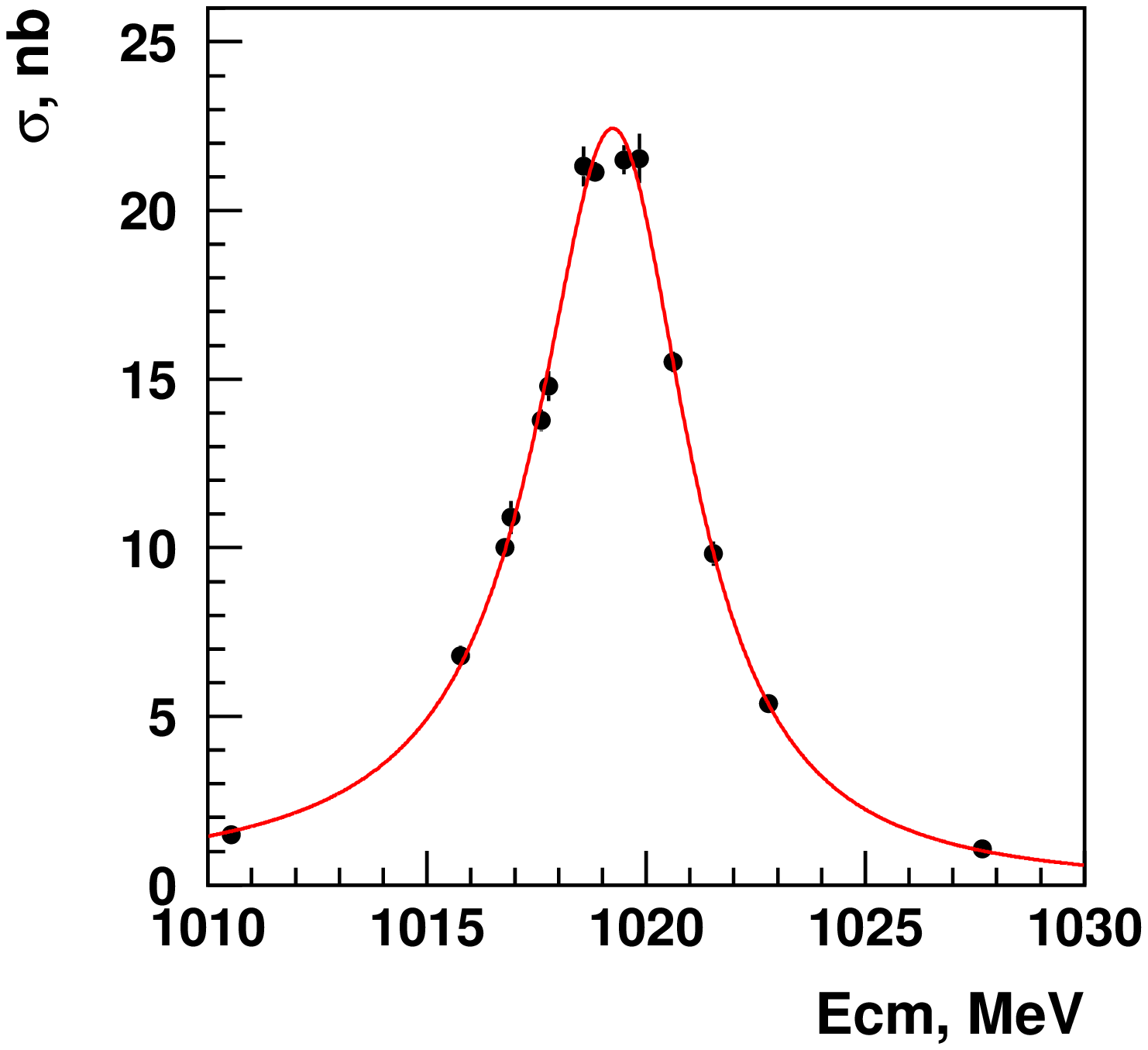}
\includegraphics[width=0.75\textwidth]
{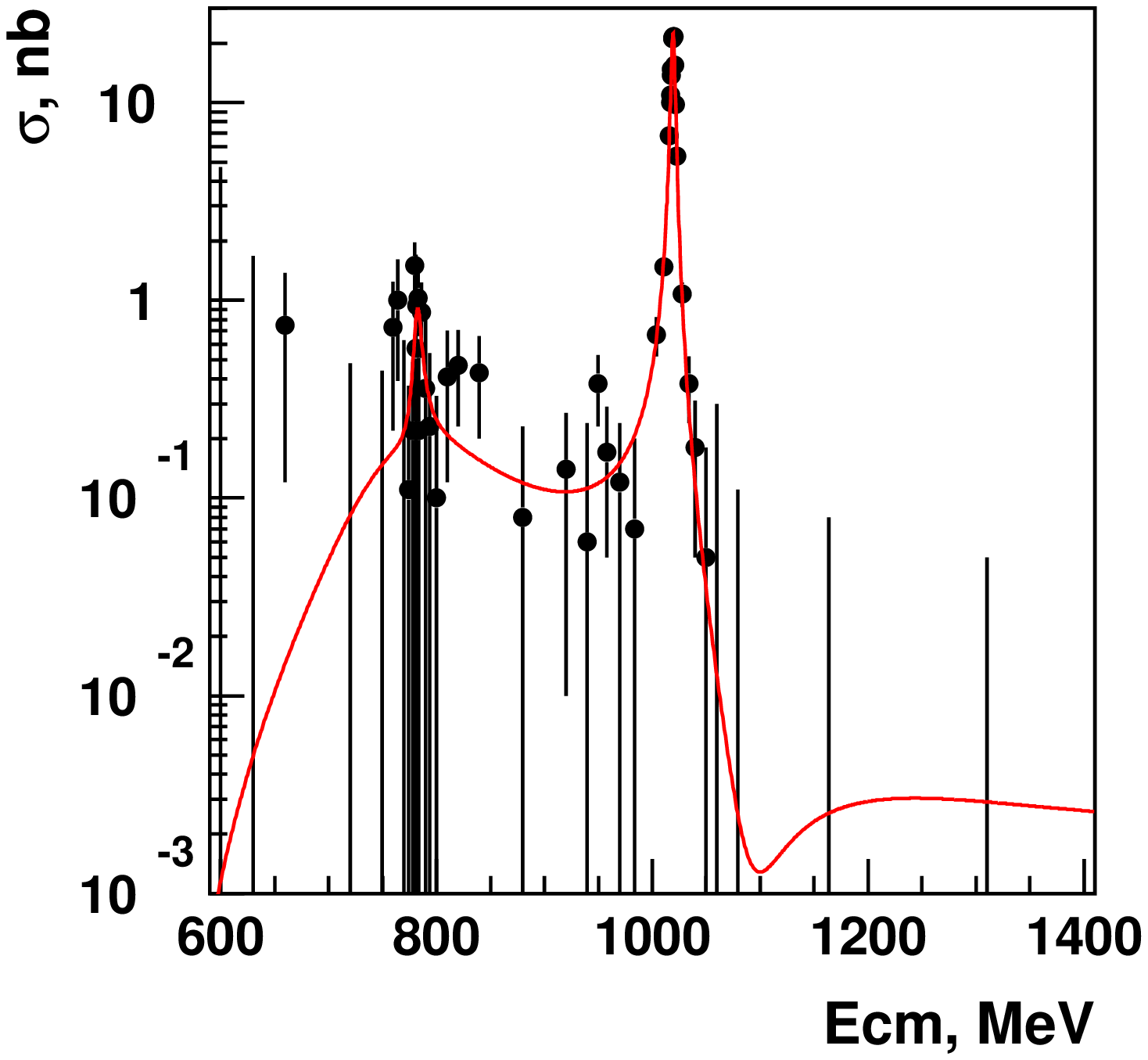}
\caption{The cross section of the process $\eeetag$. The points
  with error bars represent the experimental data, the curve
  corresponds to the result of the fit.}
  \label{fig:etag_cs}
\end{figure*}

\begin{figure*}
\centering
\includegraphics[width=0.5\textwidth]
{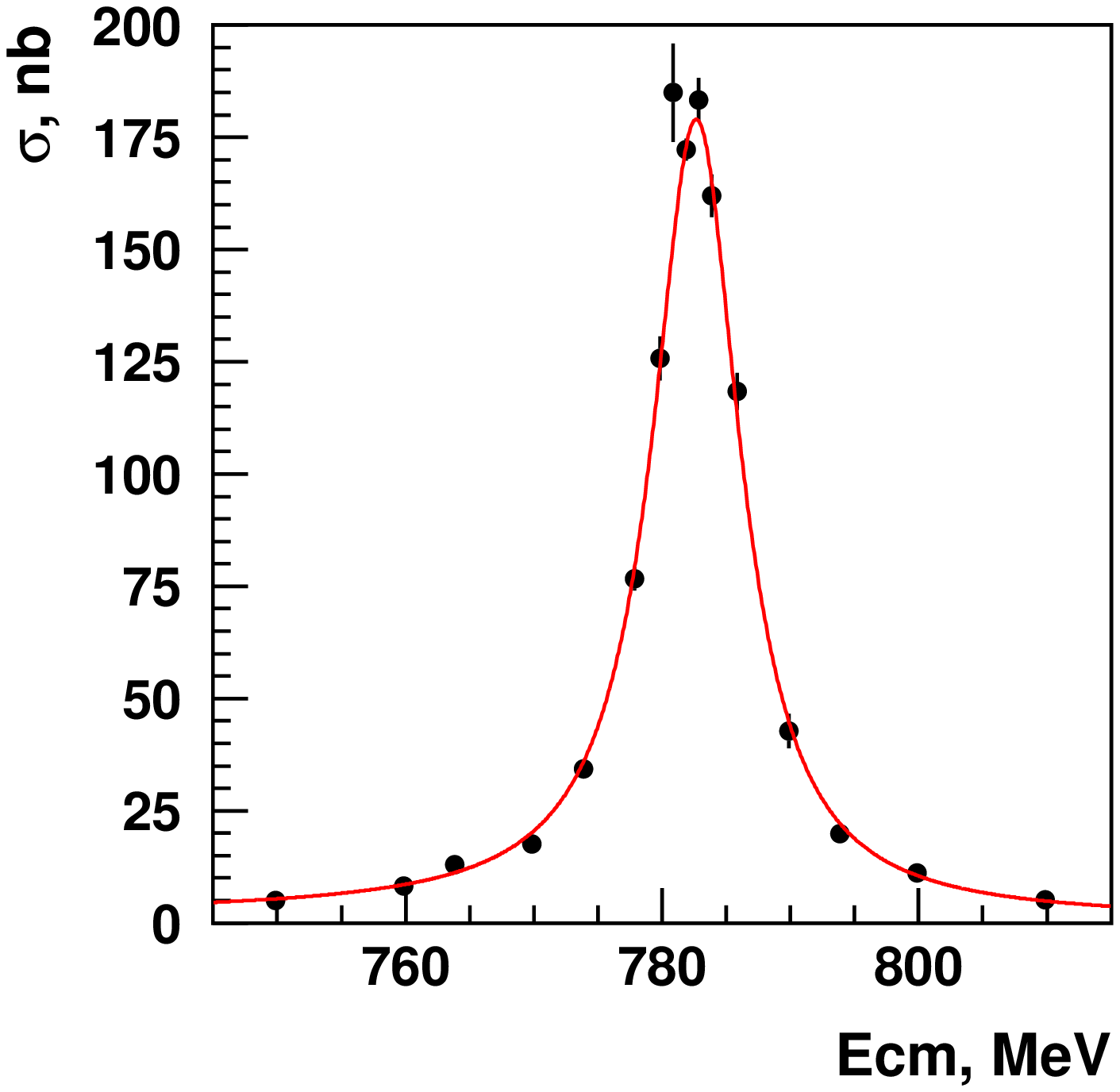}\hfill
\includegraphics[width=0.5\textwidth]
{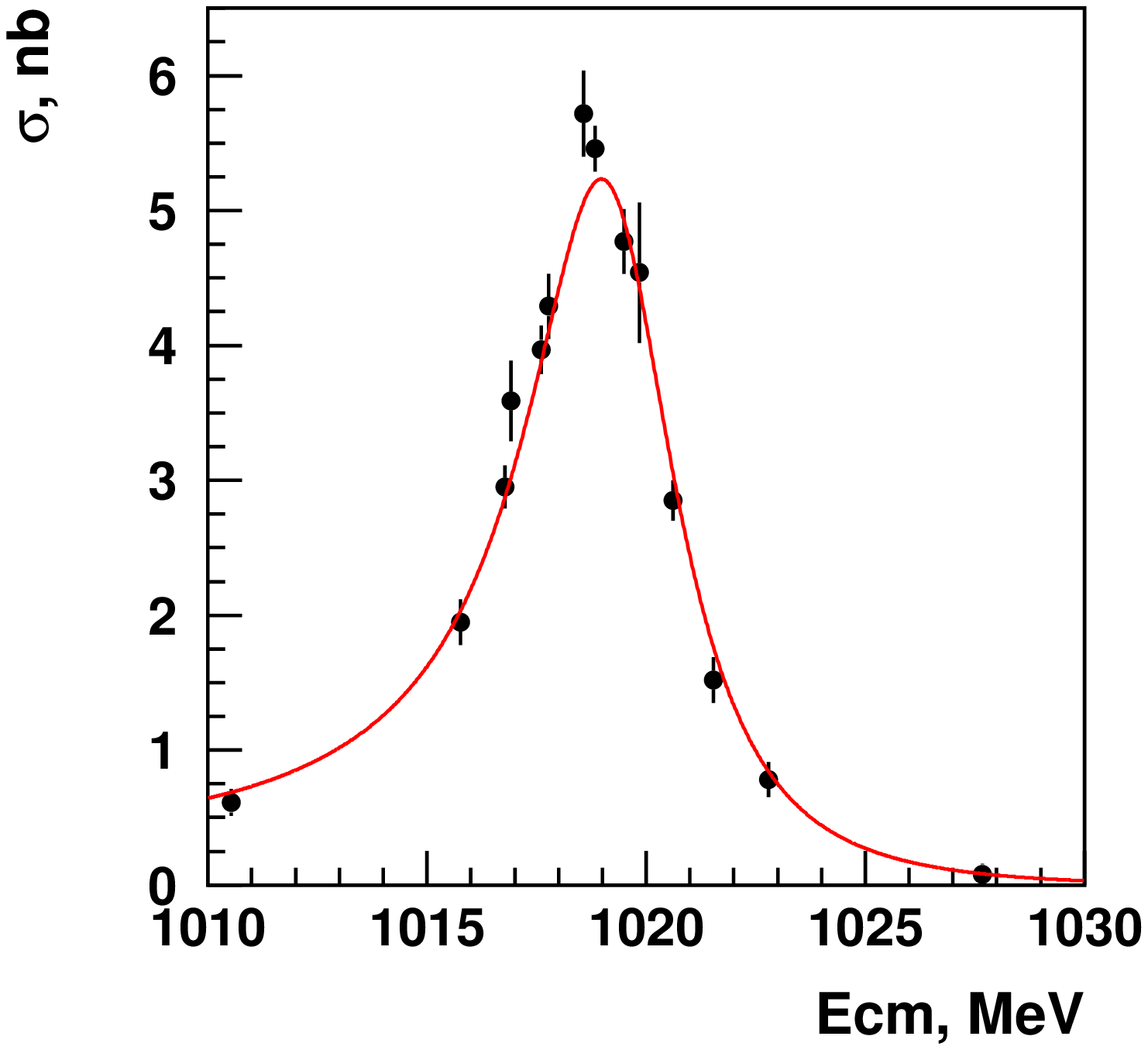}
\includegraphics[width=0.75\textwidth]
{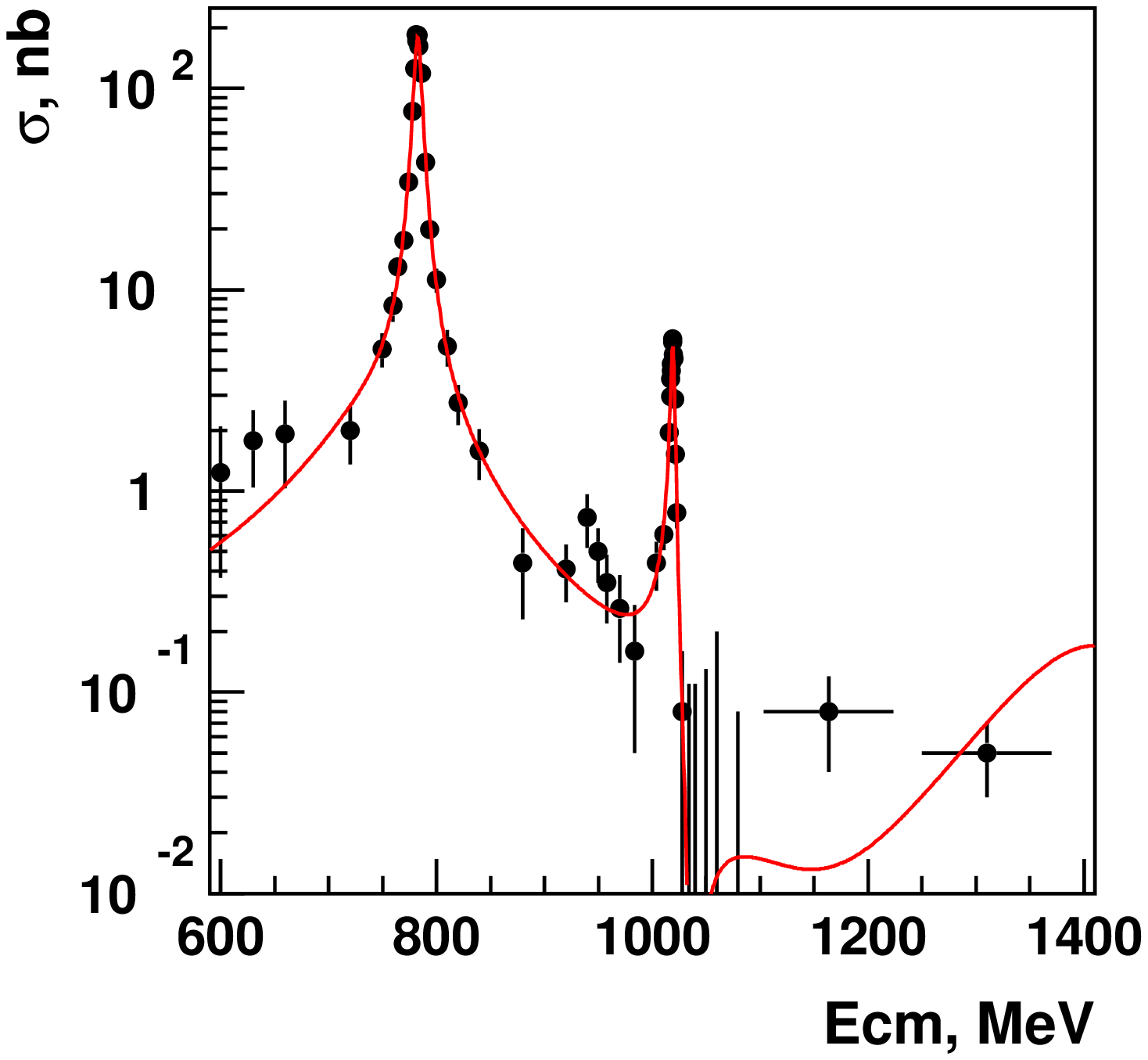}
\caption{The cross section of the process $\eepig$. The points
  with error bars represent the experimental data, the curve
  corresponds to the result of the fit.}
  \label{fig:pig_cs}
\end{figure*}

The obtained  cross sections of the processes $\eeetag$, $\pig$ 
are shown in Figs.~\ref{fig:etag_cs},~\ref{fig:pig_cs}.
 The observed pattern of the energy dependence is due to the 
 interference of the $\rho$, $\omega$ and $\phi$ mesons. 
The detailed information on this analysis is listed in
Tables~\ref{tab:etag_cs1}-\ref{tab:pig_cs2}.
The cross section shown there is a so called ``dressed'' one,
which is used in the approximation of the energy dependence 
with resonances. For applications to various dispersion integrals 
like that for the leading order hadronic contribution to the muon 
anomalous magnetic moment, the ``bare'' cross section should be 
used~\cite{rho}.

The maximum likelihood method is applied to fit the energy dependence
of the experimental cross sections obtained from the relation (\ref{Nth}).

The Born cross section of these processes can be written as:
\begin{eqnarray}
  \label{eq:cross-section}
  \sigma_{{\rm P}\gamma}(s)=\frac{F_{{\rm P}\gamma}(s)}{s^{3/2}}\cdot
  \biggl|\sum_{\rm V}A_{\rm V}\biggr|^2\; , \\
  A_{\rm V}=\sqrt{\sigma^{(0)}_{\rm V}\frac{m_{\rm V}^3}
  {F(m_{\rm V}^2)}}\cdot
  \frac{m_{\rm V}\Gamma_{\rm V} e^{i\varphi_{\rm V}}}
{m^2_{\rm V}-s-i\sqrt{s}\Gamma_{\rm V}(s)} \, ,\nonumber
\end{eqnarray}
where $m_{\rm V}$ is the mass of the resonance, $\Gamma_{\rm V}(s)$ and
$\Gamma_{\rm V}=\Gamma_{\rm V}(m_{\rm V}^2)$ are its width at the squared 
c.m.energy $s$
and at the resonance peak ($s=m_{\rm V}^2$), respectively, 
$\delta_{\rm V}$ is its relative phase,
$F(s)$ is a factor taking into account the energy
dependence of the phase space of the final state,
$F_{{\rm P}\gamma}(s) = p_{\gamma}^3 =(\sqrt{s}(1 - m_{\rm P}^2/2s))^3$,
$\sigma_{\rm V}^{(0)}$ is the cross section at the resonance peak: 
\begin{equation}
  \sigma_{\rm V}^{(0)}=\sigma_{\ee\to V\to \etag}(m_{\rm V}^2)=
  \frac{12\pi B_{\rm V \to \rm{\ee}} B_{\rm V \to \etag}}{m^2_{\rm V}}\; ,
\label{eq:sigma_G}
\end{equation}
where $B_{\rm V\to \rm{\ee}}$ and $B_{\rm V\to\etag}$ are the 
corresponding branching ratios. In Eq.~(\ref{eq:cross-section}) we sum 
over all vector mesons relevant at this energy, 
$V=\rho, \omega, \phi, \rho^{\prime}, \omega^{\prime}$.  

The Gounaris-Sakurai model has been used for the description of the
$\rho$ meson~\cite{gounaris}.
To describe the energy dependence of the  $\omega$ and $\phi$ meson
widths, their main decay modes  $\pipipi$, $\pig$ as well as
$K^0_{\rm L}K^0_{\rm S}$, $K^+K^-$, $\pipipi$ and $\etag$,
respectively, were taken into account using the same
parameterization as in~\cite{kuz}.
For the $\rho^{\prime}(1450)$ the energy dependence of the width 
assumed 60\% and 40\%  branching ratios for its decays into 
$a_1(1260)\pi$ and $\omega\pi$, respectively~\cite{root}.
Its mass and width were taken to be
$1465$~MeV and $400$~MeV, respectively~\cite{pdg}.  
The energy dependence of the $\omega^{\prime}(1420)$  width is 
calculated assuming the $\omega\prime\to\rho\pi$ decay.
Its mass and width were fixed at the world average values
of 1425~MeV and 215~MeV, respectively~\cite{pdg}.
\subsection{Results of the fits} 

For the fit to the $\ee\to\etag$ cross section the resonance cross 
sections at the peak $\sigma^{(0)}_\rho$, 
$\sigma^{(0)}_\omega$, $\sigma^{(0)}_\phi$
as well as the  $\phi$ meson mass $m_{\phi}$  are free parameters.
The $\rho$ and $\omega$ meson phases are chosen to be  
$\varphi_\rho=\varphi_\omega=0^\circ$ while
that for the $\phi$ meson is $\varphi_\phi=180^\circ$ in agreement 
with the quark model. 
The values of the other parameters are taken from Ref.~\cite{pdg}.
We also consider a model in which in addition to the parameters
described above there is a contribution tentatively referred to
as that of the $\rho^{\prime}(1450)$ meson. 
A fit in the model with the  $\rho^{\prime}(1450)$ doesn't improve
$\chi^2$ and results in the value of $\sigma_{\rho\prime}^{(0)}$ 
consistent with zero, 
$\sigma_{\rho\prime}^{(0)}= 0.001^{+0.072}_{-0.001}$~nb and
compatible with our result
in the $3\pi^0$ mode~\cite{cmdeta} 
$\sigma^{(0)}_{\rho\prime} = 0.066 \pm 0.015$~nb.
 Therefore,
for our final results for the $\etag$ decay we choose a model
where $\sigma_{\rho\prime}^{0}= 0$~nb, see Table~\ref{br_fit}. 

\begin{table*}
\caption{Results of the fits for the processes $\ee \to \etag$ 
and $\ee \to \pig$}
\begin{tabular*}{\textwidth}{l@{\extracolsep{\fill}}cc}
\hline\hline
Parameter          & $\etag$   & $\pig$  \\
\hline
$\sigma_{\rho}^{(0)}$, nb     & 0.145 $\pm$ 0.063 & 
 $0.708^{+0.146}_{-0.134}$ \\
$\sigma_{\omega}^{(0)}$, nb   & $0.299^{+0.175}_{-0.124}$ & 
$154.82^{+3.29}_{-3.24}$  \\
$\sigma_{\phi}^{(0)}$, nb     & $22.791^{+0.220}_{-0.238}$ &
 $5.30 \pm 0.16$  \\
$\sigma_{\omega\prime}^{(0)}$, nb    & -- &
 $0.139^{+0.048}_{-0.051}$    \\
$m_{\phi}$, MeV    & $1019.52 \pm 0.05$   &
 1019.46 (fixed)  \\
$m_{\omega}$, MeV & 782.59 (fixed) &  $783.20 \pm 0.13$ \\
$\varphi_{\phi-\omega}, ^{\circ}$ & 180 (fixed) & $164.4 \pm 7.9$ \\
$\chi^2/$n.d.f.    & 72.3/82   & 74.0/80 \\
\hline\hline
\end{tabular*}
\label{br_fit}
\end{table*}

For the $\ee\to\pig$ case the fit parameters are:
the cross sections at the resonance peak 
$\sigma_\rho^{(0)}$, $\sigma_\omega^{(0)}$, $\sigma_\phi^{(0)}$ and the
$\omega$ meson mass $m_{\omega}$.
The $\rho-\omega$ phase is fixed to the value $13.3^\circ$  obtained in 
our study of the process $\ee\to\pi^+\pi^-$~\cite{rho}.
The $\phi-\omega$ phase is a fit parameter.
A fit, which includes a possible $\omega\prime$ contribution,
gives the best $\chi^2$ at the value of $\sigma_{\omega\prime}^{(0)}$ 
significantly differing from zero. Results of the best fit are shown 
in the last column of Table~\ref{br_fit}.

\subsection{Systematic errors}

There are two types of systematic uncertainties on the cross 
section $\sigma^{0}_{\rm V}$: 
experimental and model uncertainties.
The main sources of experimental systematic errors are 
listed below.
The systematic error due to selection criteria is 4\% estimated 
by varying the photon energy threshold, total energy deposition, 
total momentum, and $\chi^2$.
A possible uncertainty because of the method of process separation
was estimated to be 3\% by comparing our results obtained from 
fitting the distributions of the two-photon invariant mass to those 
from Dalitz plot analysis. The latter method also allows to determine 
the cross section of the QED process $e^+e^- \to 3\gamma$ and
it appears to be consistent with the theoretical prediction~\cite{ggg}:
$\sigma(3\gamma)_{\rm exp}/\sigma(3\gamma)_{\rm th}
=0.973 \pm 0.018$.
The uncertainty in the determination of the integrated luminosity 
is 1\% and comes from the selection criteria of Bhabha events, radiative
corrections and calibrations of DC and BC. 
The error of the NT 
efficiency was estimated to be 2\% by trying various
fitting functions for energy dependence and variations of the
cluster threshold. The 1\% uncertainty of the radiative corrections comes 
from the dependence on the emitted photon energy and the accuracy of
the theoretical formulae. In total, the experimental systematic
uncertainty of the cross section  is  6\%. 

The model uncertainty estimated by comparing the values of the
cross section at the resonance peak in various models differing by
the values of phases and resonance parameters  was
1\% (2\%) for the $\rho$, 3\%(0.1\%) for the $\omega$ and 0.1\%(5\%) 
for the $\phi$ meson in the $\etag$ and $\pig$ decay modes, 
respectively.

\section{Discussion}

In Table~\ref{product} we present our results  
in terms of the product of the branching ratios 
$\br(V\to\ee)\times\br(V \to P\gamma)$, where $P=\eta(\pi^0)$,
which is calculated
from $\sigma_{\rm V}^{(0)}$ according to (\ref{eq:sigma_G}). For the
$\etag$ mode one should additionally take into account the
branching ratio of the $\eta \to \gamma \gamma$ decay taking its
value
$\br(\eta \to \gamma\gamma)= (39.43 \pm 0.26)\%$ from Ref.~\cite{pdg}.
For the $\pig$ mode the corresponding value
$\br(\pi^0 \to \gamma\gamma)= (98.798 \pm 0.032)\%$ from Ref.~\cite{pdg}
was included at the MC generation stage. Our results are in good 
agreement with the world average values~\cite{pdg}. 

\begin{table*}
\caption{ $\br(V \to \ee)\times\br(V \to P\gamma)$}
\begin{tabular*}{\textwidth}{l@{\extracolsep{\fill}}cc}
\hline\hline
  Decay                 & This work & PDG--2004 \\
\hline
$\rho\to\etag, 10^{-8}$  & $1.50 \pm 0.65 \pm 0.09$ & $1.38 \pm 0.17$ \\
$\omega\to\etag, 10^{-8}$ & $3.17^{+1.85}_{-1.31} \pm 0.21$ 
& $3.53 \pm 0.35$  \\
$\phi\to\etag, 10^{-6}$    & $4.093^{+0.040}_{-0.043} \pm 0.247$   
& $3.85 \pm 0.07$ \\
\hline
$\rho\to\pig, 10^{-8}$    & $2.90^{+0.60}_{-0.55} \pm 0.18$
& $2.8 \pm 0.6$ \\
$\omega\to\pig, 10^{-6}$  & $6.47 \pm 0.14 \pm 0.39$ 
& $6.37^{+0.17}_{-0.15}$ \\
$\phi\to\pig, 10^{-7}$ & $3.75 \pm 0.11 \pm 0.29$ & $3.67 \pm 0.28$ \\ 
\hline
\hline
\end{tabular*}
\label{product}
\end{table*}

By dividing the product of the branching ratios above by the
corresponding world average leptonic width from Ref.~\cite{pdg}
one can obtain the branching ratios of the radiative decays
confronted in Table~\ref{brad2} to the world average values~\cite{pdg}.
\begin{table*}
\caption{ $\br(V \to P\gamma)$}
\begin{tabular*}{\textwidth}{l@{\extracolsep{\fill}}cc}
\hline\hline
   Decay                & This work & PDG--2004 \\
\hline
$\rho\to\etag, 10^{-4}$    & $\brhoetag$   & $3.0 \pm 0.4$  \\
$\omega\to\etag, 10^{-4}$  & $\bomegaetag$ & $4.9 \pm 0.5$   \\
$\phi\to\etag, 10^{-2}$         & $\bphietag$   & $1.295 \pm 0.025$   \\
\hline
$\rho\to\pig, 10^{-4}$     & $\brhopig$    & $6.0 \pm 1.3$ \\
$\omega\to\pig, 10^{-2}$   & $\bomegapig$  & $8.92^{+0.28}_{-0.24}$ \\
$\phi\to\pig, 10^{-3}$     & $\bphipig$    & $1.23 \pm 0.10$ \\
\hline
\hline
\end{tabular*}
\label{brad2}
\end{table*}

Taking into account a variation of the $m_{\omega}$ and 
$m_{\phi}$ in various models as well as a systematic error
caused by the uncertainties of the beam energy 
calibration, we obtain for the resonance masses:
\begin{eqnarray}
 m_{\omega} & = & 783.20 \pm 0.13 \pm 0.16~{\rm MeV}, \\
 m_{\phi} & = & 1019.52 \pm 0.05 \pm 0.05~{\rm MeV},
\end{eqnarray}  
consistent with the world average values
$782.59 \pm 0.11$~MeV and $1019.456 \pm 0.020$~MeV, 
respectively~\cite{pdg}.


Our result for the cross section of the process $e^+e^- \to \etag$
at the peak of the $\phi$ meson can be combined with the 
independent measurement of the same quantity in the decay mode
$\eta \to 3\pi^0$ performed at CMD-2~\cite{cmdeta} to obtain the ratio
of the branching fractions of the $\eta$ meson, 
$\br(\eta\to 3\pi^0)/\br(\eta\to\gamma\gamma)$. Since in both cases
the $\eta$ meson decays into neutral particles only, most of
systematic uncertainties will cancel in such a ratio.
As a result, two sources of the systematic error survive: 2.5\%  
for the selection criteria and 3\% due to process separation and we obtain 
\begin{equation}
\label{eq:comp_etag1}
\frac{\br(\eta\to 3\pi^0)}{\br(\eta\to\gamma\gamma)}
=0.817\pm 0.012\pm 0.032,
\end{equation}
which is consistent with the world average value
$0.825 \pm 0.007$~\cite{pdg}.

\section{Conclusions}

\begin{itemize}
\item
Using a data sample corresponding to integrated luminosity of 
21~pb$^{-1}$, the cross sections of the processes $\eeetag$, $\pig$ 
have been measured in the c.m. energy range 600--1380~MeV.
The following branching ratios have been determined: 
\begin{eqnarray*}
\br(\rho^0\to\etag)&=&(\brhoetag) \cdot 10^{-4}, \\
\br(\omega\to\etag)&=&(\bomegaetag) \cdot 10^{-4},\\
\br(\phi\to\etag)&=&(\bphietag) \cdot 10^{-2}, \\
\br(\rho^0\to\pig)&=&(\brhopig) \cdot 10^{-4}, \\
\br(\omega\to\pig)&=&(\bomegapig) \cdot 10^{-2}, \\
\br(\phi\to\pig)&=&(\bphipig) \cdot 10^{-4}.
\end{eqnarray*}
\item
From the two independent measurements of the $\phi\to\etag$ decay
the following ratio of the branching ratios of the $\eta$ meson has
been obtained:  
\begin{eqnarray*}
\br(\eta\to 3\pi^0)/\br(\eta\to\gamma\gamma)
=0.817\pm 0.012\pm 0.032.
\end{eqnarray*}
\end{itemize}

{\bf Acknowledgments} 

The authors are grateful to the staff of VEPP-2M for the
excellent performance of the collider, and to all engineers and 
technicians who participated in the design, commissioning and operation
of CMD-2. This work is supported in part 
by grants DOE DEFG0291ER40646, NSF PHY-0100468,
PST.CLG.980342, RFBR-03-02-16280, RFBR-03-02-16477, RFBR-03-02-16843,
RFBR-04-02-16217, and RFBR-04-02-16223-a.


\begin{thebibliography}{99}
\bib{th1} T.~Ohshima, Phys. Rev. D 22 (1980) 707.

\bib{th2} P.J.~O'Donnell, Rev. Mod. Phys. 53 (1981) 673.

\bib{th3} A.~Bramon, A.~Grau, G.~Pancheri, Phys. Lett.
B 344 (1995) 240.

\bib{th4}
M.~Hashimoto, Phys. Rev. D 54 (1996) 5611.

\bib{th5} M.~Benayoun, S.I.~Eidelman, V.N.~Ivanchenko,
 Z. Phys. C  72 (1996) 221.

\bib{th6} P.~Ball, J.M.~Frere, M.~Tytgat,
 Phys. Lett. B 365 (1996) 367.

\bib{th7} A.~Bramon, R.~Escribano, M.D.~Scadron,
 Eur. Phys. J. C  7 (1999) 271.

\bib{th8} M.~Benayoun, 
Phys. Rev. D 59 (1999) 114027.

\bib{amm} M.~Davier, et al., Eur. Phys. J. C 31 (2003) 503.

\bib{hyb1} T.~Barnes, et al., Phys. Rev. D 55 (1997) 4157.

\bib{hyb2} F.~Close, A.~Donnachie, Yu.S.~Kalashnikova, 
Phys. Rev. D 67 (2003) 074031.

\bib{pdg}
{S.~Eidelman et al., Phys. Lett. B 592 (2004) 1.}

\bib{nd84} {V.P.~Druzhinin, et al., Phys. Lett. 144 B (1984) 136.} 

\bib{nd89} {S.I.~Dolinsky et al., Z. Phys. C42 (1989) 511.}

\bib{snd00} {M.N.~Achasov et al., Eur. Phys. J. C 12 (2000) 25.}

\bib{snd03} {M.N.~Achasov et al., Phys. Lett. B 559 (2003) 171.}


\bibitem{cmddet} {G.~A.~Aksenov et al., 
Preprint Budker INP 85-118, Novosibirsk, 1985; \\
E.~V.~Anashkin et al., ICFA Instr. Bulletin 5 (1988) 18.}

\bibitem{prep} {R.~R.~Akhmetshin et al., 
Preprint Budker INP 99-11, Novosibirsk, 1999.}

\bibitem{mccmd} {E.V.~Anashkin et al., 
Preprint Budker INP 99-1, Novosibirsk, 1999.}

\bib{cmdeta} {R.R.~Akhmetshin et al., 
Phys. Lett. B 509 (2001) 217.}

\bib{cmdk1} {R.R.~Akhmetshin et al., 
Phys. Lett. B 508 (2001) 217.}

\bib{cmdk2} {R.R.~Akhmetshin et al., 
Phys. Lett. B 551 (2003) 27.}

\bib{cmdop1} {R.R.~Akhmetshin et al., 
Phys. Lett. B 562 (2003) 173.}

\bib{cmdop2} {R.R.~Akhmetshin et al., 
Phys. Lett. B 580 (2004) 119.}

\bibitem{radcor}{E.~A.~Kuraev and V.S.~Fadin, 
Sov. J. Nucl. Phys., 41 (1985) 466.}
 
 
\bibitem{rho} {R.R.~Akhmetshin et al., 
Phys. Lett. B 578 (2004) 285.}


\bibitem{gounaris}{G.~J.~Gounaris and J.J.~Sakurai, 
Phys. Rev. Lett. 21 (1968) 244.} 

\bib{kuz} {R.R.~Akhmetshin et al., 
Phys. Lett. B 434 (1998) 426.}

\bibitem{root}
R.R.~Akhmetshin et al., Phys. Lett. B 466 (1999) 392.


\bib{ggg}
S.I.~Eidelman and E.A.~Kuraev, Nucl. Phys. B 143 (1978) 353.

\end{thebibliography}
\end{document}